\pdfoutput=1

\documentclass[11pt,prd,aps,amssymb,amsmath,tightenlines,showpacs]{revtex4}
\usepackage{graphicx}

\usepackage{hyperref} 
\hypersetup{colorlinks=true,linkcolor=black, citecolor=blue, urlcolor=black}

\begin{document}

\title{Isospectral Hermitian counterpart of complex non Hermitian\\
Hamiltonian $\ p^{2}-gx^{4}+a/x^{2}$}

\author{Asiri Nanayakkara}\email{asiri@ifs.ac.lk}
\author{ Thilagarajah Mathanaranjan${^*}$}\email{mathan@jfn.ac.lk}

\affiliation{$^*$Institute of Fundamental Studies Hanthana Road, Kandy, Sri Lanka\\
$^\dagger$Department of Mathematics, University of Jaffna, Sri Lanka}

\begin{abstract}
\bigskip
In this paper we show that the non-Hermitian Hamiltonians $H=p^{2}-gx^{4}+a/x^2$\ and the conventional Hermitian Hamiltonians $h=p^2+4gx^{4}+bx$
 ($a,b\in \mathbb{R}$) are isospectral if $a=(b^2-4g\hbar^2)/16g$ and $a\geq -\hbar^2/4$. This new class includes the equivalent non-Hermitian -Hermitian
Hamiltonian pair, $p^{2}-gx^{4}$ and $p^{2}+4gx^{4}-2\hbar \sqrt{g}x,$ found by Jones and Mateo six years ago as a special case. When $a=\left(b^{2}-4g\hbar ^{2}\right) /16g$ and $a<-\hbar^2/4,$ although $h$ and $H$ are still isospectral, $b$ is complex and $h$ is
no longer the Hermitian counterpart of $H$.

\end{abstract}


\maketitle

\section{Introduction}\label{sec:1}

Bender and Boettcher in a pioneering paper \cite{R1} showed that non Hermitian, $PT$-symmetric Hamiltonians of the form
\begin{equation}
H_{0}=p^{2}-g\left( ix\right)^{N}  \label{eq:1}
\end{equation}
posses real and positive eigenspectra when $N\geq2$. Since then many 1-D
$PT$-symmetric non Hermitian Hamiltonian models have been investigated both
quantum mechanically as well as classically. Interest in non-Hermitian $PT$-symmetric models has increased considerably during the last decade mainly
due to their usefulness in the areas such as particle-physics, quantum
optics, supersymmetric and magnetohydrodynamics and now the applicability
and the usefulness of non-Hermitian $PT$-symmetric quantum mechanics have been
well established \cite{R1,R2,R3,R4,R5,R6,R7,R8}. If $PT$-symmetry is not spontaneously broken,
Non-Hermitian $PT$-symmetric Hamiltonians have real\textbf{\ }energy spectra.
However, for a given $PT$-symmetric Hamiltonian, there is no simple way of
figuring out ahead of time whether the $PT$-symmetry is spontaneously broken
or not. Mostafazadeh \cite{R6} has proved that if the Hamiltonian of a quantum
system possesses an exact $PT$-symmetry (unbroken $PT$-symmetry) then the
Hamiltonian is equivalent to a Hermitian Hamiltonian which has the same
spectrum. This was achieved by constructing the unitary operator relating a
given non-Hermitian Hamiltonian with exact $PT$-symmetry to a Hermitian
Hamiltonian. Nonetheless, only in a few instances, people succeeded in
finding Hermitian Hamiltonians which posses the same eigenspectra as $PT$-symmetric non Hermitian Hamiltonians \cite{R7,R8,R9,R10,R11,R12}.\\

Using operator techniques and path integral methods, Jones et al \cite{R7,R8,R9} found
that the complex non-Hermitian $PT$-symmetric Hamiltonian $p^{2}-gx^{4}$ and
the conventional Hermitian Hamiltonian $p^{2}+4gx^{4}-2\sqrt{g}x$ are
isospectral$.$ However using a method based on a combination of certain
integrals (viz. Fourier) and point (i.e. change-of-variables)
spectrum-preserving transformations, Buslaev and Grecchi \cite{R10} had shown this
equivalence relation several years earlier. It is also interesting to note
that these results had been published five years earlier than the pioneering
paper \cite{R1} on $PT$-symmetry by Bender et al.\\

Recently, the Asymptotic Energy Expansion (AEE) method has been applied by
Nanayakkara et al to show that the complex non-Hermitian $PT$-symmetric
Hamiltonian $p^{2}-gx^{4}+4i\hbar \sqrt{g}x$ and the conventional Hermitian
Hamiltonian $p^{2}+4gx^{4}+6\hbar \sqrt{g}x$ have the same eigenspectra \cite{R12}.\\

In this paper we show that the Hamiltonians $\ H=p^{2}-gx^{4}+a/x^{2}$ and $%
h=p^{2}+4gx^{4}+bx$ are isospectral if $a=\left( b^{2}-4g\hbar ^{2}\right)
/16g$ and the $p^{2}-gx^{4}+4i\hbar \sqrt{g}x$ and $p^{2}+4gx^{4}+6\hbar 
\sqrt{g}x$ as well as $p^{2}-gx^{4}$ and $p^{2}+4gx^{4}-2\hbar \sqrt{g}x$
are special cases of $H$ and $h$. The outline of the paper is as follows. In
Sec. \ref{sec:2}, the equivalence condition for $H$ and $h$ is derived using the AEE method. The behavior of eigenenergies and breakdown of $PT$-symmetry
with respect to the parameters of the Hamiltonians are investigated in
Sec. \ref{sec:3}. Exact ground state wave functions, superpotentials and
supersymmetric partners of both Hamiltonians are analyzed in Sec. \ref{sec:4}. Concluding remarks are given in Sec. \ref{sec:5}.

\section{Derivation of equivalence condition}\label{sec:2}

In this section, we establish the conditions for which the non-Hermitian $PT$- symmetric quartic Hamiltonian
\begin{equation}
H=p^{2}-gx^{4}+\frac{a}{x^{2}} \label{eq:2}
\end{equation}%
and the conventional Hermitian Hamiltonian
\begin{equation}
h=p^{2}+\alpha x^{4}+bx  \label{eq:3}
\end{equation}
are equivalent. Here \ $a$, $g$, $\alpha $ and $b$ are assumed to be real.
However, later in the Sec. \ref{sec:3}, we consider the cases where these
parameters are complex as well. In a previous study on equivalent non
Hermitian and Hermitian Hamiltonians \cite{R12}, it was shown that the
Hamiltonians $p^{2}-gx^{4}+4i\hbar \sqrt{g}x$ and $p^{2}+4gx^{4}+6\hbar 
\sqrt{g}x$ are equivalent with zero energy ground states and the
supersymmetric partner of $-gx^{4}+4i\hbar \sqrt{g}x$ is $-gx^{4}+\frac{%
2\hbar ^{2}}{x^{2}}.$ Further these two Hamiltonians are found to be
isospectral as well. Consequently, the Hamiltonian $p^{2}-gx^{4}+\frac{%
2\hbar ^{2}}{x^{2}}$ is equivalent to $p^{2}+4gx^{4}+6\hbar \sqrt{g}x$ and,
hence $a=2\hbar ^{2},\alpha =4g,$ and $b=6\hbar \sqrt{g}$ are one set of
parameters for which (\ref{eq:2}) and (\ref{eq:3}) are equivalent. Therefore, it is worthwhile
to investigate whether there are any other parameter values for which $H$
and $h$ are equivalent.\\

In order to obtain the general conditions of equivalence, we used the
Asymptotic Energy Expansion (AEE) method \cite{R12,R13,R14} which is employed
by Nanayakkara et al. The AEE method is an analytic method where each term
in the expansion can be obtained explicitly in terms of Gamma functions and
multinomials of the parameters in the potential. The accuracy and the
applicability of AEE method to obtain equivalent Hamiltonians have been
demonstrated in \cite{R12}.  First the AEE is derived for the non Hermitian
Hamiltonian $H$. Since Hamiltonian $H$ contains a $1/x^{2}$ term, the
standard AEE method used for even degree polynomial potentials has to be
modified. Therefore the complete derivation is described below.\\\\\\\\\\
Consider the non Hermitian Hamiltonian $H$
\begin{equation}
H\left( x,p\right) =p^{2}+V\left( x\right) \label{eq:4}
\end{equation}
where\textbf{\ }$V\left( x\right) =-gx^{4}+\frac{a}{x^{2}}.$\\\\
The AEE\ quantization condition for this potential is
\begin{equation}
J\left( E\right) =n\hbar   \label{eq:5}
\end{equation}%
where $n$\ is a positive integer and quantum action variable $J\left(E\right) $\ is given by
\begin{equation}
J\left( E\right) =\frac{1}{2\pi }\underset{\gamma }{\int }P\left( x,E\right)dx   \label{eq:6}
\end{equation}
$P\left( x,E\right) $\textbf{\ }satisfies the Riccati equation
\begin{equation}
\frac{\hbar }{i}\frac{\partial P\left( x,E\right) }{\partial x}+P^{2}\left(
x,E\right) =E-V\left( x\right) =P_{c}\left( x,E\right)  \label{eq:7}
\end{equation}%
Note that $P\left( x,E\right) $\ relates to the wave function as $P\left(
x,E\right) =\frac{\hbar }{i}\frac{\partial \Psi /\partial x}{\Psi }.$\ The
contour $\gamma $\ in (\ref{eq:6}) encloses two physical turning points of $
P_{c}\left( x,E\right) $. Boundary conditions imposed upon $P\left(
x,E\right) $ is $P\left( x,E\right) \rightarrow $ $P_{c}\left( x,E\right) $
as $\hbar \rightarrow 0$ \cite{R15,R16}.\\\\
For the above potential, (\ref{eq:7}) becomes
\begin{equation}
\frac{\hbar }{i}\frac{\partial P\left( x,E\right) }{\partial x}+p^{2}\left(
x,E\right) =E+gx^{4}-\frac{a}{x^{2}}.  \label{eq:8}
\end{equation}
Let $\epsilon =E^{-1/4}$ and $y=g^{1/4}\epsilon x.$ Then (\ref{eq:8}) becomes, after
simplification,
\begin{equation}
\hat{h}y^{2}\epsilon ^{5}\frac{\partial P\left( y,\epsilon \right) }{%
\partial y}+y^{2}\epsilon ^{4}P^{2}\left( y,\epsilon \right)
=y^{2}(1+y^{4})-ag^{1/2}\epsilon ^{6}   \label{eq:9}
\end{equation}
where $\hat{h}=\frac{\hbar }{i}g^{1/4}$. In order to obtain asymptotic
energy expansion, first $P\left( y,\epsilon \right) $\ is expanded as an
asymptotic series in powers of $\epsilon $ and subsequently obtain
recurrence relations.\textbf{\ }This expansion usually has zero radius of
convergence.\textbf{\ }However, truncating the series after a finite number
of terms provides a good approximation to $P\left( y,\epsilon \right) $
\cite{R17,R18}. The asymptotic series expansion is written as
\begin{equation}
P\left( y,\epsilon \right) =\epsilon ^{s}\overset{\infty }{\underset{k=0}{%
\sum }}a_{k}\left( y\right) \epsilon ^{k}   \label{eq:10}
\end{equation}
where $a_{k}$ and $s$ are determined below. Substituting (\ref{eq:10}) in (\ref{eq:9}) and
equating coefficients of $\epsilon ^{0}$, $s$ and $a_{0}$ are found as $s=-2$
and $a_{0}=\sqrt{1+y^{4}}$ and (\ref{eq:9}) becomes

\begin{equation}
\hat{h}\overset{\infty }{\underset{k=0}{y^{2}\sum }}\epsilon ^{k+3}\frac{%
da_{k}}{dy}+y^{2}\underset{i=0}{\overset{\infty }{\sum }}\overset{\infty }{%
\underset{j=0}{\sum }}a_{i}a_{j}\epsilon
^{i+j}=y^{2}(1+y^{4})-ag^{1/2}\epsilon ^{6}  \label{eq:11}
\end{equation}
Next assume $a_{k}=0$ when $k<0$ and rearranging terms,
\begin{equation}
\ \left( \hat{h}\overset{\infty }{y^{2}\underset{k=1}{\sum }}\frac{da_{k-3}}{%
dy}+\underset{k=1}{\overset{\infty }{y^{2}\sum }}\overset{k-1}{\underset{i=1}%
{\sum }}a_{i}a_{k-i}+2y^{2}a_{0}\overset{\infty }{\underset{k=0}{\sum }}%
a_{k}\right) \ \epsilon ^{k}=y^{2}(1+y^{4})-ag^{1/2}\epsilon ^{6}.  \label{eq:12}
\end{equation}\\
Then coefficients $a_{k}$'s are given by

\begin{equation}
a_{k}=\frac{-1}{2y^{2}a_{0}}\left[ y^{2}\underset{i=1}{\overset{k-1}{\sum }}%
a_{i}a_{k-i}+\hat{h}y^{2}\frac{da_{k-3}}{dy}+ag^{1/2}\delta _{k,6}\right] . 
\label{eq:13}
\end{equation}
In the above formula $a_{k}=0\ \forall k<0.$ Now $J$ can be written as
\begin{equation}
J\left( E\right) =\overset{\infty }{\underset{k=0}{\sum }}b_{k}E^{\frac{%
-(k-3)}{4}} \label{eq:14}
\end{equation}
where
\begin{equation}
b_{k}=\frac{1}{2\pi }\underset{\gamma }{\int }a_{k}dy \label{eq:15}
\end{equation}
and can be determined analytically in terms of $g$ and $a.$ The contour $
\gamma $ encloses the two branch points of $\sqrt{1+y^{4}}$ (i.e. $e^{i\pi
/4}$ and $e^{3i\pi /4}$) on the complex plane. The quantization condition $%
J\left( E\right) =n\hbar $ determines the eigenenergies of $H.$\\

Using (\ref{eq:13}) and evaluating the integral  (\ref{eq:15}) analytically, the asymptotic
series is obtained. The eigenenergy expansion becomes
\begin{equation}
J\left( E\right) =\underset{k=0}{\overset{\infty }{\sum }}b_{k}E^{\frac{%
-(k-3)}{4}}.  \label{eq:16}
\end{equation}
Here first six non zero $b_{k}$'s are
\begin{equation}
b_{0}=\frac{\Gamma \left[ \frac{1}{4}\right] }{3g^{1/4}\sqrt{2\pi }%
\smallskip \ \Gamma \left[ \frac{3}{4}\right] }, \label{eq:17}
\end{equation}
\begin{equation}
b_{3}=-\frac{\hbar }{2}, \label{eq:18}
\end{equation}
\begin{equation}
b_{6}=\frac{g^{1/4}(4a-\hbar ^{2})\ \Gamma \left[ \frac{3}{4}\right] }{4%
\sqrt{2\pi }\smallskip \ \Gamma \left[ \frac{1}{4}\right] },\label{eq:19}
\end{equation}
\begin{equation}
b_{12}=\frac{g^{3/4}(80a^{2}-200ah^{2}-11\hbar ^{4})\ \Gamma \left[ \frac{1}{%
4}\right] }{1536\sqrt{2\pi }\smallskip \ \Gamma \left[ \frac{3}{4}\right] },
\label{eq:20}
\end{equation}
\begin{equation}
b_{18}=-\frac{77g^{5/4}(192a^{3}-1296a^{2}h^{2}+1860ah^{4}+61\hbar ^{6})\
\Gamma \left[ \frac{3}{4}\right] }{30720\sqrt{2\pi }\smallskip \ \Gamma %
\left[ \frac{1}{4}\right] },  \label{eq:21}
\end{equation}
\begin{equation}
b_{24}=-\frac{1105g^{7/4}(256a^{4}-3328a^{3}h^{2}+14432a^{2}h^{4}-17360a%
\hbar ^{6}+353h^{8})\ \Gamma \left[ \frac{1}{4}\right] }{3670016\sqrt{2\pi }%
\smallskip \ \Gamma \left[ \frac{3}{4}\right] }.  \label{eq:22}
\end{equation}
The next step is to obtain the AEE expansion for the Hamiltonian $h$ in (\ref{eq:3}).
Since the AEE expansion for $h$ has been derived in \cite{R12}, only the result is
presented below. The expansion of the quantum action variable $J(E)$ for the
Hamiltonian $h$ is

\begin{equation}
J^{\prime }\left( E\right) =\underset{k=0}{\overset{\infty }{\sum }}\beta
_{k}E^{\frac{-(k-3)}{4}}.\label{eq:23}
\end{equation}%
The first six non zero $\beta _{k}$'s are
\begin{equation}
\beta _{0}=\frac{\Gamma \left[ \frac{1}{4}\right] }{3\sqrt{\pi }\alpha
^{1/4}\smallskip \ \Gamma \left[ \frac{3}{4}\right] },  \label{eq:24}
\end{equation}
\begin{equation}
\beta _{3}=-\frac{\hbar }{2},  \label{eq:25}
\end{equation}
\begin{equation}
\beta _{6}=-\frac{(2\hbar ^{2}\alpha -b^{2})\ \Gamma \left[ \frac{3}{4}%
\right] }{8\sqrt{\pi }\alpha ^{3/4}\smallskip \ \Gamma \left[ \frac{1}{4}%
\right] }, \label{eq:26}
\end{equation}
\begin{equation}
\beta _{12}=\frac{(44\hbar ^{4}\alpha ^{2}-60\hbar ^{2}\alpha b^{2}+5b^{4})\
\Gamma \left[ \frac{1}{4}\right] }{6144\sqrt{\pi }\alpha ^{5/4}\smallskip \
\Gamma \left[ \frac{3}{4}\right] },  \label{eq:27}
\end{equation}
\begin{equation}
\beta _{18}=\frac{77(488\hbar ^{6}\alpha ^{3}-636\hbar ^{4}\alpha
^{2}b^{2}+90\hbar ^{2}\alpha b^{4}-3b^{6})\ \Gamma \left[ \frac{3}{4}\right] 
}{245760\sqrt{\pi }\alpha ^{7/4}\smallskip \ \Gamma \left[ \frac{1}{4}\right]
}, \label{eq:28}
\end{equation}
\begin{equation}
\beta _{24}=-\frac{1105(5648\hbar ^{8}\alpha ^{4}-6304\hbar ^{6}\alpha
^{3}b^{2}+1064\hbar ^{4}\alpha ^{2}b^{4}-56\hbar ^{2}\alpha b^{6}+b^{8})\
\Gamma \left[ \frac{1}{4}\right] }{58720256\sqrt{\pi }\alpha
^{9/4}\smallskip \ \Gamma \left[ \frac{3}{4}\right] }.  \label{eq:29}
\end{equation}\\\\
By equating the coefficients of $J(E)$ expansions of both Hamiltonians, the
conditions of the equivalence are obtained as
\begin{equation}
\alpha =4g,  \label{eq:30}
\end{equation}
\begin{equation}
a=\frac{b^{2}-4g\hbar ^{2}}{16g}. \label{eq:31}
\end{equation}\\
The condition (\ref{eq:30}) is obtained by equating terms $b_{0}$ and $\beta _{0}$ while condition (\ref{eq:31}) is derived by equating $b_{6}$ and $\beta _{6}$. When these two conditions are satisfied, it was found that $b_{k}$\ and $\beta_{k}$\ are equal for next hundred $k$\ values indicating AEE of $J(E)$\ and $J^{\prime }(E)$\ identical\textbf{. }In addition, by imposing the condition that $h$ is Hermitian, the parameters $a$ and $b$ become $b^{2}\geq 0$ and $a\geq -\frac{\hbar ^{2}}{4}$.\\

Since the AEE expansion is accurate for higher eigenvalues, we have verified
the equivalence of the Hamiltonians $h$ and $H$ for low energies by solving
the Schr\"{o}dinger equation numerically along suitable contours for various
values of parameters $a$ and $b$.\\

It is evident from the Table \ref{tab:1} and Table \ref{tab:2} that both Hamiltonians $h$ and $H$
have the same eigenspectra for first ten eigenstates. On the other hand, the
expansion of $J(E)$ is very accurate for large eigenvalues and both
Hamiltonians have the identical $J(E)$ expansions as shown above.

\begin{table}[h]
\centering
\begin{tabular}{cccc}
\hline
\textbf{n} &\hspace{0.25in}\textbf{$E_{H}$}\hspace{0.25in} &\hspace{0.5in} \textbf{$E_{h}$}\hspace{0.5in} & \textbf{$E_{J}$ } \\
\hline
0 & -2.4558329 & -2.4558327 & 1.5186675 \\ 
1 & 4.5014539 & 4.5014546 & 4.5046982 \\ 
2 & 10.931991 & 10.931992 & 10.931992 \\ 
3 & 17.793015 & 17.793016 & 17.793016 \\ 
4 & 25.238132 & 25.238134 & 25.238134 \\ 
5 &33.213971& 33.213972 & 33.213972 \\ 
6 &41.666149& 41.666150 & 41.666150 \\ 
7 & 50.549802& 50.549804 & 50.549804 \\ 
8 & 59.828456& 59.828459 & 59.828459 \\ 
9 & 69.472108 & 69.472110 & 69.472110 \\ 
10 & 79.455684 & 79.455685 & 79.455685 \\ \hline
\end{tabular}
\caption{ Verification of the equivalence of Hamiltonians $%
H=p^{2}-x^{4}+\frac{6}{x^{2}}$ and $h=p^{2}+4x^{4}+10x$ . The first ten
exact eigenenergy\emph{\ }values of $H$ and $h$ and approximate eigenvalues $%
E_{J}$ obtained by $J(E)$ expansion method are given up to eight digits.}
\label{tab:1}
\end{table}

\begin{table}[h]
\centering
\begin{tabular}{cccc}
\hline
\textbf{n} &\hspace{0.25in}\textbf{$E_{H}$}\hspace{0.25in} &\hspace{0.5in} \textbf{$E_{h}$}\hspace{0.5in} & \textbf{$E_{J}$ } \\
\hline
0 &1.8961344 & 1.8961346 & 2.4545618\\ 
1 &6.0533268 &6.0533273 & 6.0884046\\ 
2 & 11.867933 & 11.867933 & 11.866200 \\ 
3 & 18.510801 & 18.510802 & 18.510890 \\ 
4 & 25.836222 & 25.836224 & 25.836220 \\ 
5 &33.733312 &33.733314 & 33.733314\\ 
6 & 42.128813 & 42.128814 & 42.128814\\ 
7 &50.969273 & 50.969275 & 50.969275\\ 
8 & 60.213679 & 60.213680 & 60.213680\\ 
9 &69.829366 & 69.829368 & 69.829368\\ 
10 & 79.789590 & 79.789590 & 79.789590 \\ 
\hline
\end{tabular}
\caption{Verification of the equivalence of Hamiltonians $%
H=p^{2}-x^{4}-\frac{1}{2x^{2}}$ and $h=p^{2}+4x^{4}+2ix$ . The first ten
exact eigenenergy values of $H$ and $h$ and approximate eigenvalues $E_{J}$
obtained by $J(E)$ expansion method are given up to eight digits.}
\label{tab:2}
\end{table}

\section{ Behavior of eigenenergies and Hermiticity}\label{sec:3}

\begin{figure}[h]
\centering  
\includegraphics[width=10cm,height=8cm]{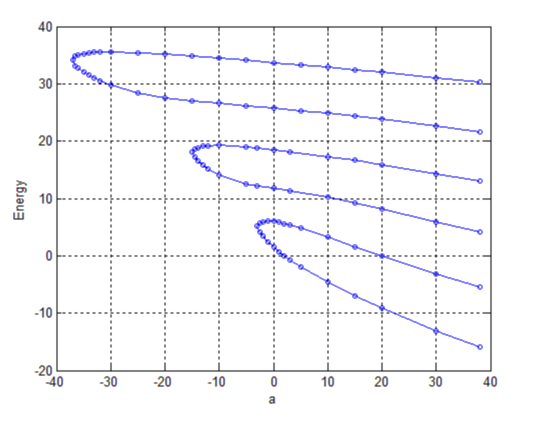}
\caption{Six lowest eigenvalues of the Hamiltonian $H$ as a function of
the parameter $a,$ when $\hbar =1$.}
\label{f:1}
\end{figure}

In the previous section, the conditions of equivalence have been
established. Next the behavior of the eigenvalues of $H$ is examined as a
function of the parameter $a$. The Hermitian condition on $h$ is relaxed
such that $b^{2}$ can also be negative. Therefore now $a$ can be less than $-%
\frac{\hbar ^{2}}{4}$ as well$.$ When $a$ is large ($\simeq 40$ and $\hbar
=1 $) lower eigenvalues of $H$ are negative as shown in figure \ref{f:1}. As $a$
decreases eigenvalues become larger and whole spectrum become real and
positive when $-2.76 < a < 2$. When $a=2$, $H$ has a zero energy ground
state and Hamiltonians $H$ and $h$ are recognized as the equivalent
Hamiltonians found by Nanayakkara et al \cite{R12}. $\ $When $a=0$, $\ b=2g\hbar$, the Hamiltonians $H$ and $h$ become the equivalent non-Hermitian -
Hermitian Hamiltonian pair found by Jones et al \cite{R7,R8}.\\

If $-\frac{\hbar ^{2}}{4}\leq a<\infty $ , $h$ is the Hermitian equivalent
Hamiltonian of the $PT$-symmetric Hamiltonian $H.$ When $a<-\frac{\hbar ^{2}}{4%
}$ , $b$ is pure imaginary and $h$ loses its Hermiticity and becomes
non-Hermitian and $PT$-symmetric. However, $h$ and $H$ are still isospectral.
Hamiltonian $h$ for this case has been studied in detail by Delabaere et al
\cite{R19} and Bender et al \cite{R21}\emph{\ }in the past. Similar to what Bender et al
have observed for the Hamiltonian $h$, as $a$ decreases below $-\frac{\hbar
^{2}}{4}$ further, adjacent pairs of energy levels of $H$ also coalesce and
then become complex, starting with the ground state and the first excited
state as shown in figure \ref{f:1}. The value of $a$ at which this coalescence takes
place for $H$ \ is $a=A=-2.76\hbar ^{2}$. Note that when $a<-\frac{\hbar ^{2}%
}{4},$ decrease in $a$ in the Hamiltonian $H$ is equivalent to an increase
in $\left\vert b\right\vert $ in $h$.

At this point it is useful to pay our attention to the Hermiticity of both
systems. We have observed previously, for $-\frac{\hbar ^{2}}{4}\leq
a<\infty ,$ $h$ is the Hermitian equivalent of $H$ and both Hamiltonians
have real spectra. $\ $When $A<a<-\frac{\hbar ^{2}}{4}$, the eigenspectrum
of $H$ is real and positive while $h$ has become non-Hermitian and $PT$-symmetric as $b$ is pure imaginary. Therefore $h$ is no longer the Hermitian
equivalent of $H.$ If the $PT$-symmetry of $H$ is not broken for $A<a<-\frac{%
\hbar ^{2}}{4}$ then by reference \cite{R6}, there exists an equivalent Hermitian
Hamiltonian which is different from $h$. However, there is another
possibility that although the eigen spectrum of $H$ is entirely real, $PT$-
symmetry of $H$ may be spontaneously broken and therefore $H$ is no longer
having a Hermitian counterpart (Note that it has not been proven that real
eigenspectra of a $PT$-symmetric system implies unbroken $PT$-symmetry) . On the
other hand when $a<A$, the lower eigenenergies of both $H$ and $h$ become
complex and hence $H$ is no longer has the true $PT$-symmetry.

\section{Unbroken Supersymmetry}\label{sec:4}

In \cite{R12}, it was shown that the Hamiltonian
\begin{equation}
H_{1}=p^{2}-x^{4}+4ix  \tag*{(32)}
\end{equation}
has zero energy ground state and the Hamiltonian
\begin{equation}
H_{2}=p^{2}-x^{4}+2/x^{2}  \tag*{(33)}
\end{equation}
is the supersymmetric partner potential (assume $\hbar =1$, $2m=1,$ and $g=1$%
). In this section we examine these two systems in detail. In a single
framework these two systems have been investigated in detail by Dorey et
al \cite{R20}. $H_{1}$ corresponds to $l=0$ and $\alpha =4$ and $H_{2}$ corresponds
to $l=1$ and $\alpha =0$ in their notations. Therefore our discussion will
be based on some of the results they have obtained in \cite{R20}. With the above
choice of $\alpha $ and $l$ , the ground state wave function of $H_{1}$ is
on the line $\alpha _{-}=0$ while the same of $H_{2}$ is on the line $\alpha
_{+}=0$ in their notations. Based on \cite{R20} and the current study, the
following results can be listed;\\\\
(1) Hamiltonians $H_{1}$ and $H_{2}$ have zero energy ground states with the
normalizable wave functions $\Phi _{0}^{(1)}(x)$ and $\Phi _{0}^{(2)}(x)$
respectively as
\begin{equation}
\Phi _{0}^{(1)}(x)=ixe^{\frac{i}{3}x^{3}}  \tag*{(34)}
\end{equation}
and
\begin{equation}
\Phi _{0}^{(2)}(x)=\left( ix\right) ^{-1}e^{-\frac{i}{3}x^{3}}  \tag*{(35)}
\end{equation}
where quantization contour starts and ends at $\left\vert
x\right\vert =\infty $ joining the \textbf{(}stokes\textbf{)} sectors $%
S_{-1} $ and $S_{1}$ and
\begin{equation}
S_{k}=\left\{ x:\left\vert \arg (x)-\frac{\pi k}{3}\right\vert <\frac{\pi }{6%
}\right\}  \tag*{(36)}
\end{equation}\\
(2) The superpotential $W_{H_{1}}\left( x\right) $ obtained from the zero
energy ground state wave function of $H_{1}$ is 
\begin{equation}
W_{H_{1}}\left( x\right) =-\frac{1+ix^{3}}{x}  \tag*{(37)}
\end{equation}\\
(3) The superpotential $W_{H_{2}}\left( x\right) $ obtained from the zero
energy ground state wave function of $H_{2}$ is 
\begin{equation}
W_{H_{2}}\left( x\right) =\frac{1+ix^{3}}{x}=-W_{H_{1}}\left( x\right) 
\tag*{(38)}
\end{equation}
(4) The supersymmetric partner Hamiltonian of $H_{1}$ is $H_{2}$ and the
supersymmetric partner Hamiltonian of $H_{2}$ is $H_{1}$ hinting at broken
supersymmetry. But both have normalizable ground state wave functions
assuring unbroken supersymmetry.\\

Therefore $H_{1}$ and $H_{2}$ are isospectral having unbroken supersymmetry
with zero energy ground states as concluded in \cite{R20}. Similar behavior has also been observed for some other systems by Znojil et al \cite{R22}.

\section{Summary and concluding remarks}\label{sec:5}

In this paper we have shown that the non-Hermitian Hamiltonian $%
H=p^{2}-gx^{4}+a/x^{2}$ is equivalent to the Hermitian Hamiltonian $%
h=p^{2}+4gx^{4}+bx$ if $a=\left( b^{2}-4g\hbar ^{2}\right) /16g$ and $a\geq -%
\frac{\hbar ^{2}}{4}.$ We applied the asymptotic energy expansion (AEE)
method to obtain the above result. The AEE method is based on series
expansion of the quantum action variables $J(E)$ in rational powers of
reciprocal of energy. The $J(E)$ expansions of these two Hamiltonians were
found to be identical. \ In addition, the spectral equivalence of $H$ and $h$
was verified with eigenspectra obtained by solving the Schr\"{o}dinger
equation for these Hamiltonians numerically along suitable contours of
integration for various values of $a$ and $b$.

When $a<-\frac{\hbar ^{2}}{4},$ it was shown that $h$ becomes non-Hermitian
and is no longer the Hermitian equivalent of $H.$ However, $H$ and $h$
remain isospectral partners even if $a<-\frac{h^{2}}{4}$. When $a$ decreases
below $a=-2.76\hbar ^{2}$, adjacent pairs of energy levels of $H$ coalesce
and then become complex conjugate pairs, starting with the ground state and
the first excited state.

\end{document}